\documentstyle[times,11pt]{article}
\newcommand{\etal}{{\it et al.\ }}

\newcommand{\vs}{{\it vs.\ }}

\newcommand{\beq}{\begin{equation}
  \renewcommand{\int}{\intop\limits}
  \renewcommand{\oint}{\ointop\limits}}
\newcommand{\eeq}{\end{equation}}
\newcommand{\beqarr}{\par\begin{minipage}{11cm} \begin{eqnarray*}}
\newcommand{\eeqarr}{\end{eqnarray*} \end{minipage} \hfill 
   \stepcounter{equation}{\rm (\theequation)}\vspace{3mm}\linebreak}
\newcommand{\bdm}{\begin{displaymath}
  \renewcommand{\int}{\intop\limits}
  \renewcommand{\oint}{\ointop\limits}}
\newcommand{\edm}{\end{displaymath}}

\newcommand{\up}[1]{\ifmmode^{\rm #1}\else$^{\rm #1}$\fi}

\newcommand{\arcd}{\ifmmode^{\circ}\else$^{\circ}$\fi}
\newcommand{\arcm}{\ifmmode{'}\else$'$\fi}
\newcommand{\arcs}{\ifmmode{''}\else$''$\fi}

\newcounter{pagefrom}
\newcounter{pageto}
\newcounter{volume}
\newcounter{year}

\newenvironment{Titlepage}{
\vspace*{2cm}
  \begin{center}
}{
  \end{center}\par\vspace{3mm}
}

\newcommand{\Title}[1]{{\large\bf\boldmath #1 \\[3mm] {\footnotesize by} 
\\[3mm]}}

\newcommand{\Author}[2]{{\large\spaceskip 2pt plus 1pt minus 1pt #1}\\[3mm]
   {\small #2}\\[6mm]}

\newcommand{\Received}[1]{}

\newcommand{\Abstract}[2]{{\footnotesize\begin{center}ABSTRACT\end{center}
\vspace{1mm}\par#1\par
\noindent
{\bf Key words:~~}{\it #2}}}

\newcommand{\FigCap}[1]{\footnotesize\par\noindent Fig.\  %
  \refstepcounter{figure}\thefigure. #1\par}

\newcommand{\TabCap}[2]{\begin{center}\parbox[t]{#1}{\begin{center}
  \small {\spaceskip 2pt plus 1pt minus 1pt T a b l e}
  \refstepcounter{table}\thetable \\[2mm]
  \footnotesize #2 \end{center}}\end{center}}

\newcommand{\TableFont}{\footnotesize}

\newcommand{\MakeTable}[4]{\begin{table}[htb]\TabCap{#2}{#3}
  \begin{center} \TableFont \begin{tabular}{#1} #4 
  \end{tabular}\end{center}\end{table}}

\newcommand{\MakeTableSep}[4]{\begin{table}[p]\TabCap{#2}{#3}
  \begin{center} \TableFont \begin{tabular}{#1} #4 
  \end{tabular}\end{center}\end{table}}

\newenvironment{references}%
{
\footnotesize \frenchspacing

\newcommand{\ApJS}{Astrophys.\ J.~Suppl.~Ser.}

\newcommand{\AJ}{Astron.\ J.}

\newcommand{\Acta}{Acta Astron.}
\newcommand{\MNRAS}{MNRAS}
\renewcommand{\and}{{\rm and }}
\section{{\rm REFERENCES}}
\sloppy \hyphenpenalty10000
\begin{list}{}{\leftmargin1cm\listparindent-1cm
\itemindent\listparindent\parsep0pt\itemsep0pt}}%
{\end{list}\vspace{2mm}}

\def\TYLDA{~}
\newlength{\DW}
\newcommand{\refitem}[5]{\item[]{#1} #2%
\def\REFARG{#3}\ifx\REFARG\TYLDA\else, {\it#3}\fi
\def\REFARG{#4}\ifx\REFARG\TYLDA\else, {\bf#4}\fi
\def\REFARG{#5}\ifx\REFARG\TYLDA\else, {#5}\fi.}

\newcommand{\Section}[1]{\section{\normalsize\bf#1}}

\newcommand{\Acknow}[1]{\par\vspace{5mm}{\bf Acknowledgements.} #1}
\newcommand{\coo}[7]{$\alpha_{#1}=#2\up{h}#3\up{m}#4\up{s}$,
$\delta_{#1}=#5\arcd#6\arcm#7\arcs$}
\newcommand{\FigurePs}[7]{\begin{figure}[htb]\vspace{#1}
\includegraphics{#4}
\FigCap{#2}\label{#3}
\end{figure}}
\newcommand{\udot}[1]{\makebox[0pt][l]{.}\up{#1}}
\begin{document}
\newfont{\bb}{timesbi at 12pt}
\def\thefootnote{\fnsymbol{footnote}}
\begin{Titlepage}
\Title{The All Sky Automated Survey. The Catalog of the Short Period Variable
Stars in the Selected Fields \footnote{Based on observations obtained 
at the Las Campanas Observatory of the Carnegie Institution of Washington.}}
\Author{G.~~P~o~j~m~a~\'n~s~k~i}{Warsaw University Observatory
Al~Ujazdowskie~4, 00-478~Warsaw, Poland\\
e-mail: gp@sirius.astrouw.edu.pl}

\end{Titlepage}

\vspace*{-12pt}
\Abstract{Results of the first two month of observations using the All Sky
Automated Survey prototype camera are presented. More than 45 000 stars
in 24 Selected Fields covering 140 sq. degrees were monitored a few times
a night resulting in the $I$-band catalog containing $10^7$
individual measurements. Period search revealed 126 periodic variables
brighter than 13 mag. Only 30 of them are known variables included in GCVS.
The other 90 objects are newly detected variables - mainly eclipsing binaries
(75\%) and pulsating stars (17\%). We estimate that completeness of the current
catalogs of variable stars is smaller than 50 \% already for the stars brighter
than 9 mag.  The Catalog is accessible over the WWW:
{\em http://www.astrouw.edu.pl/$\sim$gp/asas/asas.html}
}{Catalogs -- Stars:variables -- Surveys} 

\vspace*{-6pt} 
\Section{Introduction}
The All Sky Automated Survey (Pojma{\'n}ski 1997, hereafter Paper I) 
is a new observing project 
which ultimate goal is detection and investigation of any kind of the 
photometric variability present all over the sky (Paczy{\'n}ski 1997). 
We want to achieve this aim at relatively low cost, using simple  
automatic modules. In 1997 we have started monitoring over 20 Selected Fields to
the limiting magnitude 13 (in $I$-band) using prototype automated mount
equipped with 768 $\times$ 512 MEADE Pictor~416 CCD camera, 135~mm
f/1.8 telephoto lens and $I$-band (Schott RG-9, 3mm) filter. The instrument
was placed at the Las Campanas Observatory which is operated by
the Carnegie Institution of Washington, in the vicinity of the new OGLE-2
telescope (Udalski, Kubiak and Szyma{\'n}ski 1997), where room for
the control computer was kindly allocated.

During the routine observations control program loops over the list of
selected fields pointing camera and taking 3 minute exposures. 
Dark and flat-field images are exposed at the beginning of the night and
appropriate data reduction process is applied after data acquisition in
the fully automated way.
Results of the aperture photometry are put into the ASAS Catalog, from
which they may be retrieved e.g. using the World Wide Web.
Detailed description of the prototype instrument, data acquisition and 
reduction process and ASAS Catalog can be found in Paper I.

This paper presents results of the search for periodic variables in
the ASAS Catalog using data obtained during the first two month of the
prototype instrument operation. 

\Section{Observations, Data Reduction and  Period Search}

Twenty four 2 $\times$ 3 deg fields were selected for the test run at 
the Las Campanas Observatory and for the follow-up observations. To
obtain the largest possible diversity of images required for tests we have 
selected some  standard calibration fields (e.g. PG1323-086, S-107 from
Landolt (1992)), crowded fields (e.g. Milky Way in Centaurus, LMC),
star-poor regions (e.g. Octans), ecliptic fields (e.g. Aquarius) and
galaxy rich area (in Virgo). Coordinates of the Selected Field centers
as well as number of exposures $N_{exp}$, number of measured stars $N_{star}$,
number of detected
variables $N_{var}$ and detection rate $f_{det} = N_{var}/N_{star}$
are given in Table 1.

\MakeTable{|l|r|r|r|r|r|r|}{10cm}{\label{table1}
Selected Fields observed during the first two month of ASAS operation}{
\hline
\multicolumn{1}{|c|}{Field~ID} & \multicolumn{1}{c|}{$\alpha_{2000}$} & \multicolumn{1}{c|}{$\delta_{2000}$} & \multicolumn{1}{c|}{$N_{exp}$} & \multicolumn{1}{c|}{$N_{star}$} & \multicolumn{1}{c|}{$N_{var}$} & \multicolumn{1}{c|}{$f_{det}$}\\
&\multicolumn{1}{c|}{hh:mm}&\multicolumn{1}{c|}{dd:mm}&&&&\\
\hline
LMC~1 & 05:10 & -68:10 & 85 & 949 & 3 & 0.0032\\
LMC~2 & 05:10 & -70:00 & 86 & 1037 & 2 & 0.0019\\
LMC~3 & 05:40 & -68:10 & 92 & 1109 & 3 & 0.0027\\
LMC~4 & 05:40 & -70:00 & 94 & 1159 & 1 & 0.0008\\
Centaurus~1 & 11:35 & -60:00 & 279 & 6497 & 19 & 0.0029\\
Centaurus~2 & 11:35 & -61:50 & 268 & 5676 & 33 & 0.0058\\
Centaurus~3 & 11:35 & -63:40 & 262 & 5358 & 25 & 0.0045\\
Octans~1 & 12:00 & -85:00 & 328 & 2035 & 5 & 0.0025\\
Virgo & 12:30 & 03:00 & 156 & 505 & 1 & 0.0019\\
Coal~Sack & 12:50 & -63:00 & 294 & 5147 & 16 & 0.0031\\
PG1323-086 & 13:25 & -08:50 & 192 & 641 & 0 & ~\\
Centaurus~4 & 13:50 & -30:00 & 211 & 1255 & 4 & 0.0032\\
Centaurus~5 & 13:50 & -31:50 & 206 & 1511 & 0 & ~\\
Libra & 15:05 & -15:00 & 178 & 1074 & 2 & 0.0018\\
S-107 & 15:40 & -00:20 & 83 & 682 & 0 & ~\\
Sagittarius~1 & 17:00 & -22:30 & 74 & 2524 & 1 & 0.0003\\
Sagittarius~2 & 18:00 & -23:30 & 68 & 3262 & 4 & 0.0010\\
Sagittarius~3 & 19:00 & -22:30 & 71 & 2987 & 0 & ~\\
Corona~Australis & 19:00 & -40:00 & 79 & 2185 & 3 & 0.0014\\
Sagittarius~4 & 20:00 & -20:30 & 71 & 1246 & 2 & 0.0016\\
Capricornus~1 & 21:00 & -17:00 & 66 & 715 & 1 & 0.0013\\
S-113 & 	21:40 & -00:20 & 59 & 563 & 1 & 0.0018\\
Capricornus~2 & 22:00 & -12:00 & 58 & 448 & 0 & ~\\
Aquarius & 	23:00 & -06:30 & 44 & 265 & 0 & ~\\
\hline
\multicolumn{3}{|c|}{Totals} & 3404 & 46244 & 126 & 0.0027\\
\hline
}

Even though Selected Fields were observed only if the air-mass of the 
frame center was smaller than 2 and the angular distance from the Moon was
larger than 45 deg, we have encountered some
problems with flat correcting the frames (cf. aper I). 
This resulted in some offset (up to 0\udot{m}1)
between magnitudes of the same stars measured in the
overlapping areas of the Centaurus and LMC fields. The data from 
the overlapping fields were processed and stored separately.

Photometric calibration was performed
using Landolt (1992) standard fields: PG131323-086, S-107 and S-113.
Since our project is monochromatic at present we were not able to use
color terms in the magnitude calibration - we have determined only the
zero point $i_{\rm 0}$ of the  transformation  between instrumental $i$ 
and standard $I_{\rm cat}$ magnitudes:
\beq
\label{eq1}
I_{cat} = i_{\rm 0} + i + k_{\rm I}X.
\eeq
where $k_{\rm I}$ and $X$ denote extinction coefficient in the $I$-band and 
air mass, respectively.

For the photometric nights of April 28 and 29 with
average $k_{\rm I}$ extinction of 0.058 and 0.047 respectively
we have found that for 20 stars in the magnitude range $8 < I_{cat} < 12$
zero point of the transformation amounts to:
$i_{\rm 0} = 0.34 \pm 0.035 $.
One must be aware that lacking color terms and having problems
with flat correction  our $I_{cat}$ values might
be in some cases erroneous even by a tenth of a magnitude.

Transformation given by Eq. 1 was used only once for each
Selected Field - at the time it was added to the catalog for the first time.
For the remaining images median difference between
catalog and instrumental magnitudes was calculated 
and used for transformation. This was possible, since differential 
extinction over our field of view does not exceed 0\udot{m}01.
When adding a new frame to the catalog the rms difference $q$ between
the catalog magnitudes and transformed magnitudes for all stars on the
frame was calculated and stored, since it is a good indicator of the 
individual frame quality.

\FigurePs{6.5cm}
{Standard deviation $\sigma_I$ of the stellar magnitudes 
{\em vs.} $I$-band magnitudes in the 
ASAS Catalog. Black dots denote detected variables. 
Most of the remaining points
with large $\sigma_I$  values are due to the long term variables.
}
{fig1}{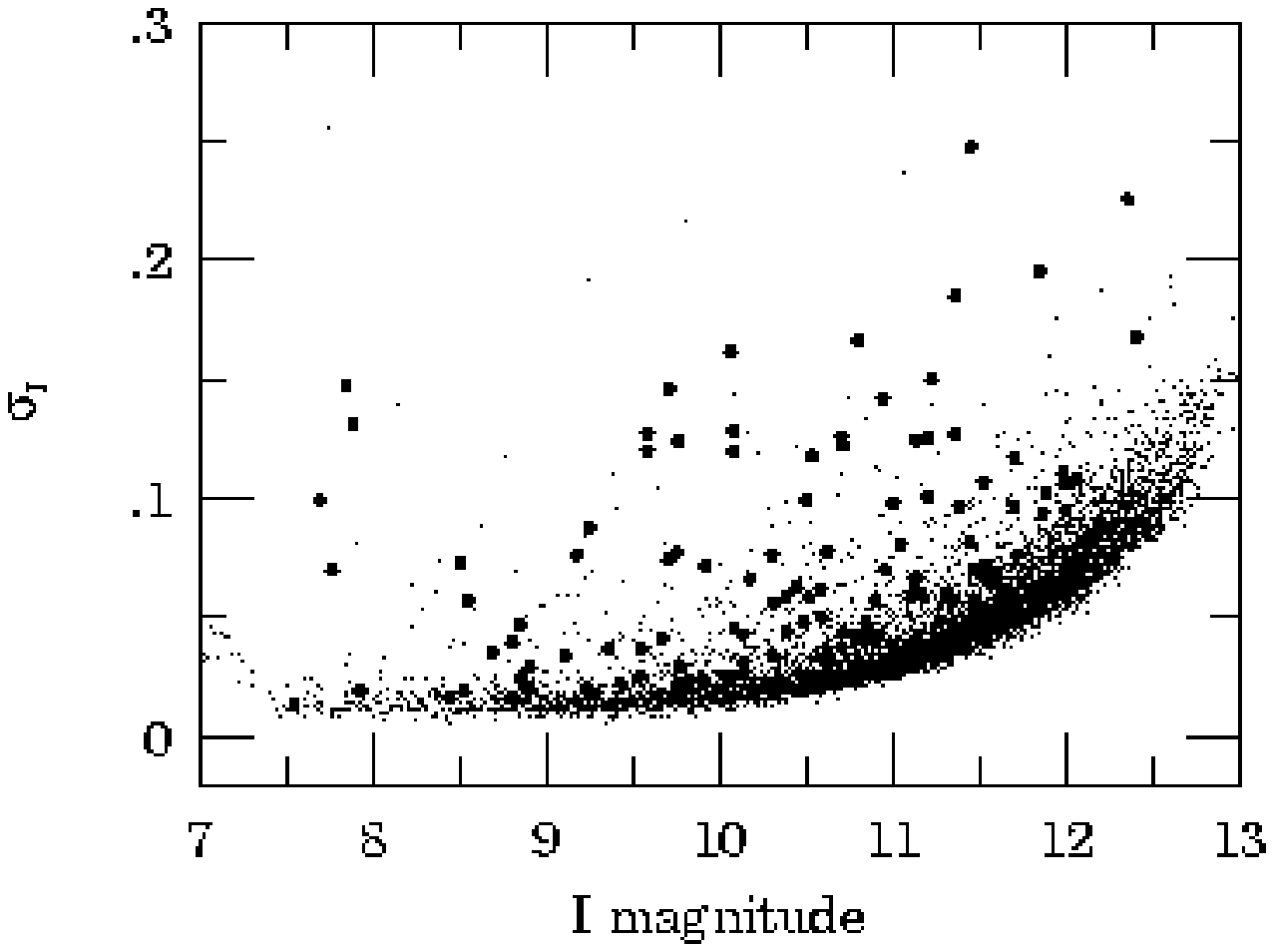}{67}{16}{-30}

The following algorithm was applied
to find periodic variable stars in the ASAS catalog:
First, the standard deviations $\sigma_I$ 
of stellar magnitudes were plotted against
the $I$-band magnitudes (Fig. fig1), and the lower envelope of the 
plotted points was determined.
Next, the stars with at least 40 measurements were selected and their
light curves extracted from the Catalog - rejecting data
coming from the frames of bad quality ($q > 0.04$).
Stellar magnitude dispersion was then
calculated and if it was large  (2 times
the envelope value) the star was tested using the analysis of variance
(AoV) method (Schwarzenberg-Czerny 1989) which proved efficient in the
OGLE variable star search (Udalski \etal 1994). 

Light curves of the stars showing large AoV statistics (detection thresold
was set to 10 - more than required for 99.5\% confidence) were then
plotted against the phase and inspected visually. 

Slowly or irregulary varying stars, which tend to produce spurious 
AoV statistics enhancements at $(k \times 0.5)^{-1} {\rm d}^{-1}$ frequencies,
as well as stars with large $\sigma_I$ deviations were stored
separately for future analysis. In few cases of low amplitude variables 
(e.g. ASAS 112552-6232.9, ASAS 113858-6328.7) we may not exclude spurious 
period asignment due to the problems with photometry quality.

\Section{The Catalog of the Short Period Variable Stars}

Stars, that passed visual inspection are presented in Table 2
containing ASAS designation, equatorial coordinates $\alpha,\delta$ 
(equinox 2000),
$I_{\rm max}$ magnitude at maximum brightness, light curve amplitude 
$\Delta I$,
epoch $T_0$ of minimum brightness (maximum for pulsating stars), period
$P$ and variability classification. Their corresponding light curves are 
collected in the Appendix.

\setcounter{table}{1}
\MakeTableSep{|r|r|r|r|r|l|r|l|}{10cm}{\label{table2}
Short Period Variable Stars in the Selected Fields.}{
\hline
\multicolumn{1}{|c|}{ID} & \multicolumn{1}{c|}{$\alpha_{2000}$} & \multicolumn{1}{c|}{$\delta_{2000}$} & \multicolumn{1}{c|}{$I_{\rm max}$} & \multicolumn{1}{c|}{$\Delta I$} & \multicolumn{1}{c|}{$T_0$} & \multicolumn{1}{c|}{$P$} & \multicolumn{1}{c|}{Type}\\
&&&&&-2450000&\multicolumn{1}{c|}{[days]}&\\
\hline
045913-6935.7 & 04:59:13.3 & -69:35:44 & 10.87 & 0.27 & 558.6047 & 0.3291 & RRc\\
050048-7029.8 & 05:00:48.0 & -70:29:52 & 10.57 & 0.37 & 558.5631 & 0.3873 & EW\\
050526-6743.2 & 05:05:27.0 & -67:43:13 & 9.78 & 0.44 & 559.563 & 4.064 & EB\\
050554-6810.8 & 05:05:54.9 & -68:10:49 & 11.22 & 0.29 & 559.114 & 1.757 & EB\\
051833-6813.5 & 05:18:32.4 & -68:13:33 & 9.94 & 0.58 & 558.5128 & 0.2854 & EW\\
052558-7011.1 & 05:25:58.1 & -70:11:08 & 10.45 & 0.17 & 557.017 & 3.168 & EB\\
052927-6852.0 & 05:29:27.4 & -68:52:02 & 8.87 & 0.06 & 558.294 & 5.216 & EB\\
054000-6828.6 & 05:39:59.7 & -68:28:41 & 10.61 & 0.52 & 558.5662 & 0.3622 & EW\\
055122-6812.7 & 05:51:21.7 & -68:12:45 & 11.53 & 0.41 & 558.9188 & 0.6434 & EW\\
110234-8530.2 & 11:02:33.2 & -85:30:13 & 11.26 & 0.17 & 553.267 & 8.381 & MISC\\
112229-6313.7 & 11:22:29.5 & -63:13:44 & 11.84 & 0.42 & 559.387 & 3.910 & EA\\
112238-6332.9 & 11:22:37.8 & -63:32:53 & 8.64 & 0.10 & 559.277 & 2.093 & ACV\\
112301-6146.8 & 11:23:01.6 & -61:46:53 & 10.14 & 0.09 & 546.331 & 2.677 & EA\\
112325-6238.7 & 11:23:25.8 & -62:38:40 & 10.13 & 0.20 & 556.767 & 4.667 & EA\\
112331-6423.2 & 11:23:30.6 & -64:23:15 & 10.88 & 0.39 & 558.2904 & 0.9056 & EA\\
112405-6403.1 & 11:24:04.8 & -64:03:05 & 11.08 & 0.52 & 558.863 & 1.703 & EB\\
112419-6344.0 & 11:24:19.0 & -63:44:00 & 11.83 & 0.33 & 559.79 & 16.15 & DCEPS\\
112502-6251.4 & 11:25:01.8 & -62:51:26 & 11.35 & 0.18 & 558.707 & 1.373 & DCEPS\\
112506-6044.1 & 11:25:05.7 & -60:44:03 & 7.53 & 0.37 & 559.817 & 5.311 & DCEP\\
112513-6122.2 & 11:25:12.9 & -61:22:08 & 7.66 & 0.20 & 548.489 & 3.216 & DCEPS\\
112552-6232.9 & 11:25:52.3 & -62:32:52 & 11.05 & 0.20 & 558.1833 & 0.7206 & EW\\
112612-6210.2 & 11:26:12.4 & -62:10:12 & 10.45 & 0.45 & 545.815 & 1.316 & EA\\
112629-6148.9 & 11:26:28.4 & -61:48:54 & 10.68 & 0.09 & 546.8368 & 0.9282 & ?MISC\\
112644-6251.8 & 11:26:44.5 & -62:51:45 & 11.12 & 0.46 & 554.908 & 1.595 & EA\\
112653-6211.9 & 11:26:53.5 & -62:11:52 & 11.70 & 0.62 & 546.6698 & 0.5877 & EW\\
112706-6037.3 & 11:27:05.9 & -60:37:19 & 8.84 & 0.22 & 559.21 & 10.57 & EA \\
112742-6127.8 & 11:27:42.7 & -61:27:53 & 11.29 & 0.22 & 546.7606 & 0.8978 & EW\\
112746-6110.5 & 11:27:46.7 & -61:10:26 & 11.48 & 0.23 & 557.357 & 1.608 & EB\\
112803-6124.7 & 11:28:03.1 & -61:24:42 & 8.47 & 0.52 & 546.239 & 3.488 & EB\\
112810-6010.5 & 11:28:10.4 & -60:10:27 & 9.50 & 0.27 & 566.41 & 19.14 & EB\\
112843-5925.7 & 11:28:42.8 & -59:25:42 & 9.58 & 0.22 & 558.1728 & 0.9220 & RRab\\
112852-6255.8 & 11:28:51.9 & -62:55:51 & 8.81 & 0.16 & 558.559 & 2.513 & EA\\
112901-6052.8 & 11:29:00.5 & -60:52:50 & 9.52 & 0.41 & 558.017 & 1.626 & EA\\
112923-6154.5 & 11:29:23.7 & -61:54:31 & 11.15 & 0.31 & 546.380 & 1.608 & EA\\
112926-6201.9 & 11:29:26.5 & -62:01:56 & 10.56 & 0.47 & 546.474 & 3.223 & EA\\
112939-5953.7 & 11:29:39.2 & -59:53:41 & 11.47 & 0.46 & 559.447 & 1.915 & EA\\
112943-6323.2 & 11:29:42.9 & -63:23:13 & 9.66 & 0.23 & 559.1225 & 0.9451 & EW\\
113019-6316.6 & 11:30:19.1 & -63:16:34 & 10.39 & 0.24 & 559.384 & 1.201 & EB\\
113023-6226.2 & 11:30:22.9 & -62:26:14 & 11.35 & 0.20 & 550.060 & 6.330 & EB\\
113043-6305.0 & 11:30:43.7 & -63:05:00 & 10.76 & 0.13 & 558.982 & 1.830 & EB\\
113131-6233.2 & 11:31:30.6 & -62:33:12 & 9.06 & 0.15 & 557.51 & 10.36 & EB\\
113149-5918.0 & 11:31:49.1 & -59:18:03 & 10.81 & 0.17 & 558.933 & 3.109 & EB\\
113210-5948.9 & 11:32:10.3 & -59:48:51 & 10.23 & 0.23 & 558.7552 & 0.4534 & EW\\
113318-6306.2 & 11:33:17.9 & -63:06:15 & 9.21 & 0.06 & 558.7505 & 0.1684 & DSCT\\
\hline
}
\setcounter{table}{1}
\MakeTableSep{|r|r|r|r|r|l|r|l|}{10cm}{
Continued}{
\hline
\multicolumn{1}{|c|}{ID} & \multicolumn{1}{c|}{$\alpha_{2000}$} & \multicolumn{1}{c|}{$\delta_{2000}$} & \multicolumn{1}{c|}{$I_{\rm max}$} & \multicolumn{1}{c|}{$\Delta I$} & \multicolumn{1}{c|}{$T_0$} & \multicolumn{1}{c|}{$P$} & \multicolumn{1}{c|}{Type}\\
&&&&&-2450000&\multicolumn{1}{c|}{[days]}&\\
\hline
113321-5949.6 & 11:33:21.4 & -59:49:32 & 11.97 & 0.42 & 558.897 & 1.134 & EB\\
113333-6353.7 & 11:33:33.4 & -63:53:41 & 9.15 & 0.28 & 558.684 & 1.982 & EA\\
113426-6320.0 & 11:34:26.2 & -63:20:02 & 10.84 & 0.12 & 558.224 & 1.718 & EA\\
113451-6128.0 & 11:34:50.9 & -61:27:57 & 10.02 & 0.14 & 547.784 & 3.422 & EB\\
113547-6104.8 & 11:35:47.7 & -61:04:45 & 10.90 & 0.14 & 557.005 & 1.122 & MISC\\
113617-6128.0 & 11:36:16.8 & -61:28:02 & 8.48 & 1.14 & 547.704 & 3.697 & EB\\
113648-6029.6 & 11:36:47.6 & -60:29:35 & 9.50 & 0.11 & 562.040 & 5.346 & EB\\
113648-6425.6 & 11:36:47.6 & -64:25:38 & 9.91 & 0.17 & 558.817 & 2.116 & EA\\
113708-6148.1 & 11:37:07.5 & -61:48:04 & 9.96 & 0.38 & 547.140 & 1.227 & EB\\
113713-5952.6 & 11:37:13.7 & -59:52:35 & 10.76 & 0.14 & 558.840 & 1.365 & EB\\
113745-6014.5 & 11:37:45.1 & -60:14:37 & 11.63 & 0.40 & 559.327 & 2.940 & EA\\
113840-6117.8 & 11:38:40.0 & -61:17:53 & 11.24 & 0.19 & 589.626 & 2.388 & EB\\
113858-6328.7 & 11:38:58.0 & -63:28:41 & 10.05 & 0.04 & 558.0529 & 0.8709 & ?MISC\\
113915-6026.1 & 11:39:15.4 & -60:26:10 & 10.32 & 0.18 & 558.4585 & 1.7472 & EB\\
113958-6449.0 & 11:39:58.1 & -64:48:59 & 9.64 & 0.37 & 556.9345 & 0.4457 & EW\\
114035-6306.2 & 11:40:35.8 & -63:06:14 & 11.78 & 0.27 & 558.9338 & 0.2859 & EW\\
114059-6241.5 & 11:40:58.6 & -62:41:33 & 7.63 & 0.47 & 558.929 & 3.347 & CEP(B)\\
114101-6036.8 & 11:41:01.4 & -60:36:48 & 8.51 & 0.04 & 554.010 & 4.411 & MISC\\
114119-6215.9 & 11:41:19.3 & -62:15:54 & 10.38 & 0.34 & 547.870 & 5.846 & DCEP\\
114141-6236.5 & 11:41:41.4 & -62:36:26 & 10.59 & 0.07 & 558.1389 & 0.6582 & MISC\\
114155-6347.7 & 11:41:54.8 & -63:47:44 & 7.91 & 0.04 & 557.806 & 1.413 & MISC\\
114201-6140.3 & 11:42:01.0 & -61:40:24 & 9.42 & 0.28 & 552.579 & 9.571 & EB\\
114248-5859.6 & 11:42:48.0 & -58:59:36 & 9.49 & 0.51 & 555.36 & 12.11 & DCEP\\
114250-6226.0 & 11:42:49.8 & -62:26:05 & 8.44 & 0.26 & 554.442 & 6.295 & EA\\
114257-6248.4 & 11:42:56.5 & -62:48:24 & 9.83 & 0.07 & 558.8346 & 0.8924 & EA\\
114308-6029.1 & 11:43:07.9 & -60:29:05 & 11.07 & 0.20 & 558.992 & 1.549 & EA\\
114345-6144.6 & 11:43:45.7 & -61:44:35 & 9.64 & 0.76 & 546.601 & 2.992 & EB\\
114353-6024.8 & 11:43:52.9 & -60:24:50 & 10.10 & 0.43 & 559.020 & 3.240 & EB\\
114357-6322.8 & 11:43:57.2 & -63:22:49 & 8.87 & 0.07 & 558.3115 & 1.715 & MISC\\
114416-6143.0 & 11:44:15.7 & -61:43:00 & 9.75 & 0.31 & 545.736 & 3.689 & EB\\
114417-6233.8 & 11:44:16.6 & -62:33:47 & 8.77 & 0.08 & 554.09 & 11.22 & DCEP\\
114510-6058.1 & 11:45:09.7 & -60:58:11 & 10.11 & 0.55 & 558.066 & 3.914 & EB\\
114555-5922.7 & 11:45:54.6 & -59:22:42 & 10.86 & 0.86 & 558.9086 & 0.4529 & RRab\\
114557-6352.9 & 11:45:57.0 & -63:52:53 & 11.69 & 0.32 & 559.0727 & 0.9538 & EA\\
114617-6100.0 & 11:46:16.6 & -61:00:06 & 10.34 & 0.15 & 558.475 & 1.227 & EB\\
114659-6228.4 & 11:46:59.1 & -62:28:25 & 11.26 & 0.64 & 558.8235 & 0.9110 & EW\\
114720-6155.1 & 11:47:19.7 & -61:55:02 & 10.57 & 0.40 & 559.054 & 1.250 & EB\\
114736-6310.4 & 11:47:36.2 & -63:10:26 & 9.85 & 0.22 & 558.8002 & 0.3393 & EW\\
114736-6322.7 & 11:47:35.9 & -63:22:39 & 10.26 & 0.25 & 558.188 & 2.314 & EB\\
114757-6225.3 & 11:47:57.7 & -62:25:15 & 8.62 & 1.23 & 559.459 & 1.657 & EB\\
114758-6034.0 & 11:47:57.3 & -60:33:58 & 12.17 & 0.50 & 560.4544 & 0.3952 & EW\\
114806-6221.3 & 11:48:05.7 & -62:21:17 & 11.50 & 0.54 & 560.154 & 4.131 & EA\\
115935-8545.9 & 11:59:36.0 & -85:45:54 & 11.08 & 0.31 & 558.6769 & 0.6111 & EW\\
122418+0351.6 & 12:24:18.5 & +03:51:34 & 12.21 & 0.63 & 555.7581 & 0.3545 & EW\\
\hline
}
\setcounter{table}{1}
\MakeTable{|r|r|r|r|r|l|r|l|}{10cm}{
Concluded}{
\hline
\multicolumn{1}{|c|}{ID} & \multicolumn{1}{c|}{$\alpha_{2000}$} & \multicolumn{1}{c|}{$\delta_{2000}$} & \multicolumn{1}{c|}{$I_{\rm max}$} & \multicolumn{1}{c|}{$\Delta I$} & \multicolumn{1}{c|}{$T_0$} & \multicolumn{1}{c|}{$P$} & \multicolumn{1}{c|}{Type}\\
&&&&&-2450000&\multicolumn{1}{c|}{[days]}&\\
\hline
123748-6219.4 & 12:37:47.6 & -62:19:22 & 9.33 & 0.11 & 558.8874 & 0.4349 & EW\\
123808-6353.8 & 12:38:07.6 & -63:53:48 & 11.49 & 1.00 & 558.144 & 1.126 & EA\\
123824-6404.8 & 12:38:24.0 & -64:04:49 & 10.27 & 0.10 & 555.849 & 7.782 & EB\\
123826-6303.9 & 12:38:26.5 & -63:03:51 & 11.94 & 0.34 & 559.373 & 1.867 & EB\\
124203-6226.2 & 12:42:02.7 & -62:26:12 & 10.87 & 0.64 & 559.013 & 1.909 & EB\\
124220-6259.6 & 12:42:20.4 & -62:59:38 & 9.11 & 0.23 & 559.649 & 1.885 & EB\\
124421-6300.8 & 12:44:21.7 & -63:00:44 & 11.09 & 0.60 & 555.05 & 12.70 & DCEP\\
124435-6331.8 & 12:44:35.2 & -63:31:46 & 9.51 & 0.08 & 559.471 & 2.570 & EB\\
125210-6312.7 & 12:52:10.0 & -63:12:40 & 10.23 & 0.22 & 555.660 & 7.937 & EA\\
125319-6401.4 & 12:53:19.0 & -64:01:23 & 10.37 & 0.20 & 558.8440 & 0.4262 & EW\\
125427-6356.1 & 12:54:27.2 & -63:56:06 & 9.61 & 0.72 & 567.06 & 23.56 & DCEP\\
125815-6258.1 & 12:58:15.1 & -62:58:08 & 11.07 & 0.52 & 557.625 & 2.526 & EA\\
125933-6210.5 & 12:59:32.8 & -62:10:36 & 11.20 & 0.21 & 559.366 & 1.544 & EB\\
125953-6159.5 & 12:59:53.4 & -61:59:33 & 10.61 & 0.37 & 558.056 & 1.482 & EB\\
130221-6328.4 & 13:02:20.5 & -63:28:23 & 10.40 & 0.09 & 558.9958 & 0.3785 & EW\\
130308-6349.4 & 13:03:07.7 & -63:49:22 & 10.10 & 0.06 & 559.836 & 5.362 & DCEPS\\
130807-8503.5 & 13:08:04.2 & -85:03:30 & 11.06 & 0.14 & 558.6656 & 0.5138 & RRc\\
131312-8528.6 & 13:13:10.5 & -85:28:34 & 10.93 & 0.29 & 558.9488 & 0.5528 & EW\\
134460-3019.2 & 13:44:59.5 & -30:19:17 & 9.00 & 0.02 & 558.2425 & 0.8734 & MISC\\
135335-2934.8 & 13:53:35.5 & -29:34:49 & 11.62 & 0.60 & 558.2297 & 0.6366 & RRab\\
135340-3036.0 & 13:53:40.1 & -30:35:59 & 8.89 & 0.16 & 558.4830 & 0.4760 & EW\\
135546-2911.5 & 13:55:46.5 & -29:11:25 & 10.61 & 0.08 & 558.5863 & 0.1476 & DSCT\\
145960-1417.0 & 15:00:00.0 & -14:17:03 & 12.14 & 0.41 & 558.8451 & 0.1675 & RRab\\
150414-1517.9 & 15:04:14.4 & -15:17:56 & 9.28 & 0.09 & 558.279 & 1.470 & EB\\
170511-2133.4 & 17:05:10.7 & -21:33:23 & 10.08 & 0.61 & 558.575 & 2.217 & EA\\
180059-2301.9 & 18:00:58.6 & -23:01:54 & 9.18 & 0.34 & 560.802 & 4.668 & EB\\
180254-2409.8 & 18:02:53.8 & -24:09:46 & 9.82 & 0.44 & 559.597 & 2.113 & EB\\
180305-2251.9 & 18:03:04.7 & -22:51:52 & 9.38 & 0.86 & 557.476 & 3.911 & EB\\
180325-2237.1 & 18:03:24.9 & -22:37:02 & 8.10 & 0.56 & 558.746 & 1.393 & EB\\
185622-4040.6 & 18:56:22.1 & -40:40:39 & 10.58 & 0.15 & 558.324 & 1.492 & EA\\
185627-4044.0 & 18:56:27.1 & -40:43:58 & 11.76 & 0.38 & 558.7765 & 0.3602 & EW\\
190537-3904.6 & 19:05:37.0 & -39:04:32 & 11.99 & 0.88 & 558.7585 & 0.3383 & EW\\
195723-2105.2 & 19:57:22.5 & -21:05:14 & 11.12 & 0.31 & 559.0770 & 0.4586 & EW\\
200338-1956.0 & 20:03:38.3 & -19:55:59 & 10.98 & 0.30 & 558.9758 & 0.9134 & EW\\
210553-1647.8 & 21:05:53.4 & -16:47:46 & 11.65 & 0.32 & 558.8694 & 0.3034 & EW\\
214610-0106.8 & 21:46:10.1 & -01:06:47 & 11.71 & 0.68 & 558.7722 & 0.2850 & EW\\
\hline
}

ASAS designation consists of the letters ASAS followed by the object
equatorial coordinates in the format hhmmss-ddmm.m. Accuracy of such
designation is consistent with the Catalog resolution of
15 arcsec, resulting from the 14.2 arcsec pixels of the prototype
instrument. Actual coordinates of the stars were calculated for individual
frames,
using astrometric solution based on star positions from the
HST Guide Star Catalog (Lasker 1988). An rms error of
astrometric transformation was usually smaller than 3 arcsec.
For about 70 variables observed in the overlapping areas of Centaurus
and LMC fields it was possible to determine independent coordinates,
based on different GSC star sets. An average rms 
difference between them was
found to be $\Delta \alpha =0\udot{s}2\pm 0\udot{s}13$ and 
 $\Delta \delta = 0\udot{"}17 \pm 1\udot{"}55$, so the $\alpha$ and $\delta$
values given in the Table 2 are presented with 1 arcsec
accuracy.

Period accuracy depends on the stellar brightness, light curve shape,
nuber of the available data points and observation time-base. Because of the data
reduction automation we decided to present periods in homogeneous format
- with the last digit usually not being very significant.

Epoch of the minimum/maximum light was determined from the phased light
curves, and its accuracy is between 0.005 and 0.05 of the period,
depending on the light curve shape.  

Light curves were classified according to the Kholopov \etal (1985) criteria
used in the GCVS catalog.
Most of the variables are EB, EW, EA, DCEP and RR stars.
Objects showing low amplitude, sinus-like variation were 
were designated MISC. A few of them (marked with ?) migth be spurious 
detections due to the bad photometry.
We did not attempt to recognize RS CVn stars yet, 
although in some cases (e.g. ASAS 112706-6037.3) such classification 
might be relevant. 

\MakeTable{|r|l|l|r||r|l|l|r|}{10cm}{\label{table3}
Known variable stars detected by ASAS.}{
\hline
\multicolumn{1}{|c|}{ASAS~ID} & \multicolumn{1}{c|}{Other~ID} &
\multicolumn{1}{c|}{Type} & \multicolumn{1}{c||}{Offset} & \multicolumn{1}{|c|}{ASAS~ID} & \multicolumn{1}{c|}{Other~ID} & \multicolumn{1}{c|}{Type} & \multicolumn{1}{c|}{Offset}\\
&&&\multicolumn{1}{c||}{["]}&&&&\multicolumn{1}{c|}{["]}\\
\hline
045913-6935.7 & XX~Dor & VAR & 31.7 & 114248-5859.6 & KK~Cen & DCEP & 6.0\\  
051833-6813.5 & RW~Dor & EW & 12.6& 114250-6226.0 & V346~Cen & EA & 2.9\\  
052558-7011.1 & \multicolumn{3}{l||}{SHV~0526293-701248}& 114345-6144.6 & MP~Cen & EB  & 3.6\\  
                 & & VAR  & 45.6 &  114416-6143.0 & MQ~Cen & EA   & 7.1\\  
112301-6146.8 & ?~V440~Cen & EA	& 50.0& 114510-6058.1 & MR~Cen & EB   & 6.0\\  
112506-6044.1 & AY~Cen & DCEP & 0.0& 114555-5922.7 & BI~Cen & RRAB & 25.8\\ 
112513-6122.2 & AZ~Cen & DCEP & 7.1& 114659-6228.4 & \multicolumn{2}{l|}{CKS91~11445-6211}& \\ 
112653-6211.9 & V343~Cen & EB & 6.9&	         & & VAR  & 4.7\\ 
112803-6124.7 & MN~Cen & EA & 2.1& 114720-6155.1 & HD~309079 & EA   & 8.3\\  
112810-6010.5 & IV~Cen & EA & 30.9& 114736-6310.4 & HD~309036 & VAR  & 2.7\\  
112901-6052.8 & LT~Cen & EA & 0.0& 114736-6322.7 & \multicolumn{2}{l|}{CKS91~11451-6253}& \\ 
113318-6314.5 & LV~Cen & DCEP & 14.7&                & &VAR  & 4.1\\      
113321-5949.6 & IW~Cen & EB & 9.6& 114757-6225.3 & SV~Cen & EB   & 10.0\\ 
113333-6353.7 & HD~100530 & DBLE & 6.6&	 114806-6221.3 & KT~Cen & EA   & 9.1\\  
113617-6128.0 & BF~Cen & EA & 5.6& 123808-6353.8 & VZ~Cru & EA   & 24.8\\ 
113958-6449.0 & TV~Mus & EW & 51.6& 125427-6356.1 & RY~Cru & VAR  & 20.6\\ 
114059-6241.5 & UZ~Cen & CEP & 4.7& 135335-2934.8 & FY~Hya & RRAB & 27.3\\ 
114119-6215.9 & IZ~Cen & DCEP & 13.8& 180059-2301.9 & WY~Sgr & EA   & 6.2\\  
114155-6347.7 & V915~Cen & ACV & 2.3& 180305-2251.9 & V792~Sgr & EB & 0.0\\  
114201-6140.3 & MO~Cen & EA & 6.3&  180325-2237.1 & V4202~Sgr & E & 6.4\\
\hline
}

\MakeTable{|l|l|r|r|r|r|r|}{10cm}{\label{table4}
GCVS variables that were not detected during ASAS variability search.}{
\hline
\multicolumn{1}{|c|}{GCVS~ID} & \multicolumn{1}{c|}{Type} & \multicolumn{1}{c|}{$\alpha_{2000}$} & \multicolumn{1}{c|}{$\delta_{2000}$} & \multicolumn{1}{c|}{$m$} & \multicolumn{1}{c|}{$\Delta~m$} & \multicolumn{1}{c|}{$P$}\\
\hline
S~Dor & SDOR & 05:18:14.0 & -69:14:59 & 8.60 & 2.90 & 0.3638\\
AA~Dor & EA & 05:31:40.5 & -69:53:10 & 11.13 & 0.47 & 0.2615\\
V384~Cen & EA & 11:39:19.7 & -62:10:20 & 11.80 & 0.60 & 12.6352\\
BG~Cen & E & 11:37:20.5 & -64:02:19 & 11.80 & 0.80 & 0.7431\\
BH~Cen & EB & 11:39:10.0 & -63:25:13 & 10.03 & 1.13 & 0.7916\\
LW~Cen & EB & 11:37:32.1 & -63:20:49 & 8.90 & 0.75 & 1.0026\\
BX~Cru & DCEP & 12:50:36.5 & -63:04:19 & 12.20 & 0.46 & 19.537\\
\hline
}

All variable stars in our survey were searched for in the GCVS catalog
and, if missing, in the SIMBAD database. 
Table 3 lists 37 matches with objects classified as variable.
30 of them are GCVS variables, 7 other are SIMBAD objects
marked as VAR. One match (V 440 Cen) is doubtful, since the GCVS object is
fainter, slightly offseted and has different period 
(3\udot{d}13 \vs 2\udot{d}67). A few
other stars show large coordinate offsets ($\geq 15$ arcsec) but are easily
identified by their properties (IV~Cen, LV~Cen, TV~Mus, IZ~Cen,
BI~Cen, VZ~Cru, RY~Cru, and FY~Hya).

The opposite test, search in the GCVS for variables located in the Selected
Fields and fulfilling selection criteria ($P < 20^{\rm d}$, $7^{\rm m}
< m < 13^{\rm m}$ and $N_{obs} > 40$), revealed 7 objects that were
missed by ASAS variability search. They are listed in Table
4; each such case was individualy inspected. \\
S~Dor is eruptive variable and was constant during
observations. \\
LW~Cen was rejected because its period (1\udot{d}002)
was too close to one day. \\
BG~Cen (\coo{GCVS}{11}{37}{20.5}{-64}{02}{19}) coordinates point close to
the blended Catalog pair (2 pixel separation)
(\coo{}{11}{37}{19}{-64}{03}{36} and \coo{}{11}{37}{22}{-64}{02}{32}).
Its period (0\udot{d}7431) can be recognized only in the second
component of the the blend, but at very low AoV statistics level.\\
BH~Cen is located 57 arcsec (4 pixels) from the bright ($I$=7\udot{m}9)
companion and its photometry is inaccurate.\\
V384~Cen is a $12^{\rm d}$ Algol system. Only 10 observations (out of
260) were obtained during eclipse, so the star was not tested for periodicity.\\
AA~Dor was missed because of its low magnitude dispersion ($\sigma_I = 0.05$).\\
BX~Cru true period ($~\sim 39^{\rm d}$) appeared to be twice the GCVS value -
too long for the present study.

Having only a small sample of missed stars we are not able to make definite
statements about completeness of the Catalog - most omitted stars are
probably blended objects and long period Algols, as well as stars with
small number of observations. We estimate that over 200 data points are
necessary to guarantee period detection for pulsating, EW and EB  stars,
and much more for Algol variables. We will address the question of
completeness in future - comparing current results and data obtained during 
the next months of ASAS operation.

\Section{Summary}

\MakeTable{|c|r|r|r|r|r|}{10cm}{\label{table5}
Number of ASAS variable stars {\em vs.} previously known variables in 1 mag bins
and completeness of the former and current data.}{
\hline
Mag & $N_{ASAS}$ & $N_{known}$ & $\frac{N_{known}}{N_{ASAS}}$ & $N_{obs}$ & 
$\frac{N_{ASAS}}{N_{obs}} \rule[-7pt]{0mm}{18pt}$\\
\hline
7-8   &  4~~~~ &  4~~~~ & 1.00~~~ &   438 & 0.009~ \\
8-9   & 13~~~~ &  5~~~~ & 0.38~~~ &  1106 & 0.012~ \\
9-10  & 28~~~~ & 12~~~~ & 0.43~~~ &  3254 & 0.009~ \\
10-11 & 42~~~~ &  9~~~~ & 0.20~~~ &  8582 & 0.006~ \\
11-12 & 36~~~~ &  6~~~~ & 0.17~~~ & 17516 & 0.002~ \\
12-13 &  3~~~~ &  0~~~~ & 0.00~~~ &  6678 & 0.0004 \\
\hline
}

The Catalog of the Short Period Variable Stars in the Selected Fields
contains 126 objects found among 45000  stars. About 90 of them where not
previously known to be variable. Results presented in Table 5
suggest that completeness of the existing catalogs is smaller than 50\%
already for stars brighter than 9 mag and 
drops significantly for fainter objects. This drop is even more
pronounced if one takes into account, that current ASAS survey starts to
be incomplete at about 11~mag.

There are many more other variable stars in the Selected Fields. Many of them
are long period variables, which were already discovered, but their period
has not been determined yet. We will perform 
detailed analysis of such objects using the data 
collected during almost one year of the
ASAS prototype instrument operation.

Our preliminary results show, that the small-scale instruments
are ideal tools for reducing incompleteness of our knowledge about 
bright objects on the sky. We are going to extend our survey
for much larger area of the sky, trying to increase the
ASAS Catalog completeness to 13~mag using longer focal length.

The Catlog of the Short Period Variable Stars, as well as the ASAS Catalog are
accessible over the World Wide Web:\\
\centerline{\em http://www.astrouw.edu.pl/$\sim$gp/asas/variables.html}\\
or\\
\centerline{\em http://www.astrouw.edu.pl/$\sim$gp/asas/asas.html}.

\vspace*{-6pt} 
\Acknow{It is a great pleasure to thank Prof.\ Bohdan Paczy{\'n}ski for 
initiative in this project, valuable discussions and providing necessary funds.

We are indebted to the OGLE collaboration for letting us use 
facilities of the Warsaw telescope and for the permanent support 
of our instrumentation
(opening and closing enclosure, tape exchanging and many more).

Special thanks are due to Andrzej Udalski for his invaluable
technical help and reanimating our equipment after serious break-down,
to Micha{\l} Szyma{\'n}ski for lots of computing hints and to
Marcin Kubiak and Przemek Wo{\'z}niak for instrument re-adjustment
after the earthquake.

I am very indebted to Carnegie Institution of Washington for providing the
excellent site for my instrument.

This work was partly supported by the KBN BST grant.}

\vbox{
\begin{center}
\Large\bf Appendix\\
Light Curves of the Short Period Variable Stars in the Selected Fields 
of the ASAS Catalog\\
(on the next pages)
\end{center}
}
\begin{figure}\vspace*{29 cm}
\includegraphics{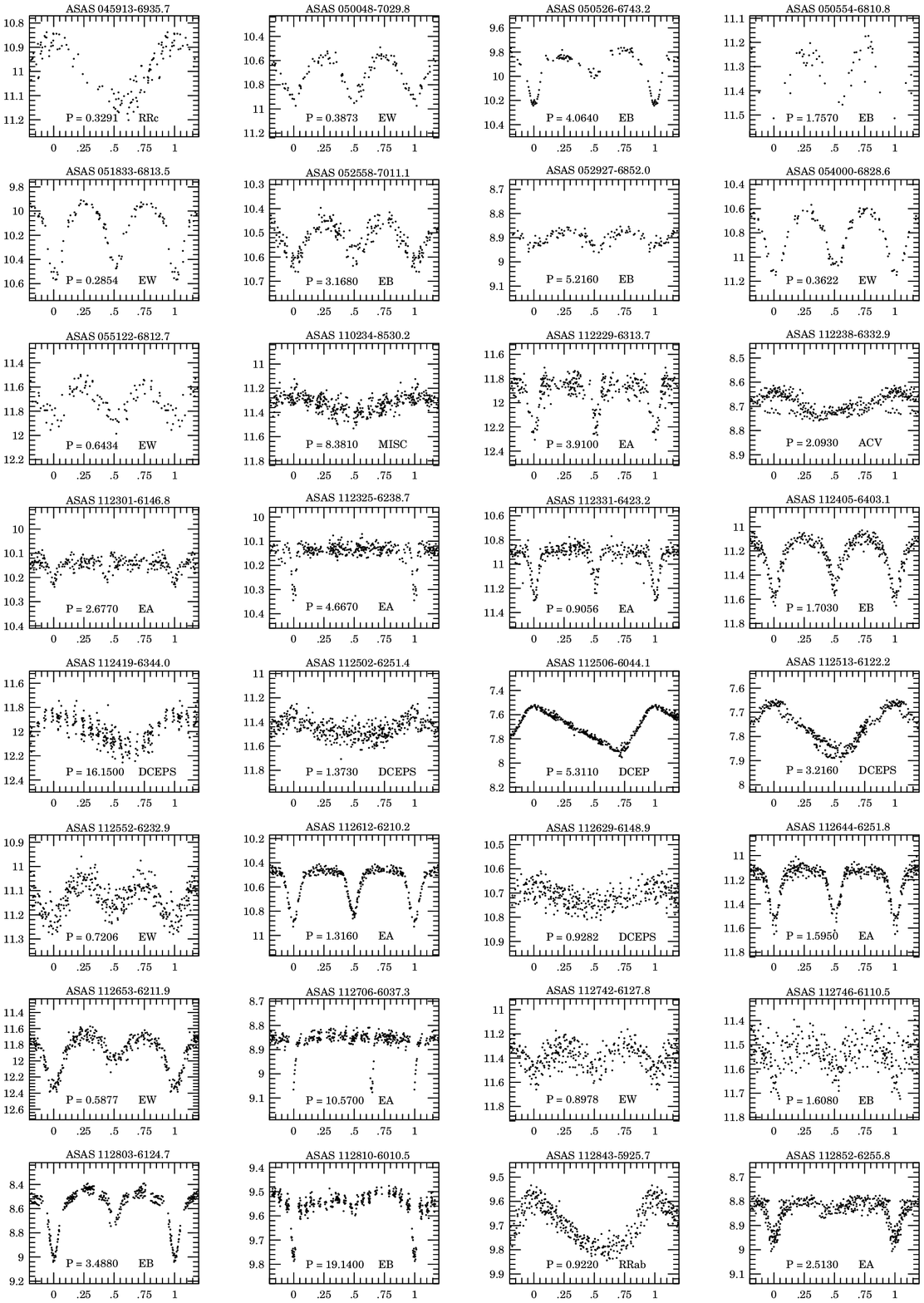}
\end{figure}
\begin{figure}\vspace*{29 cm}
\includegraphics{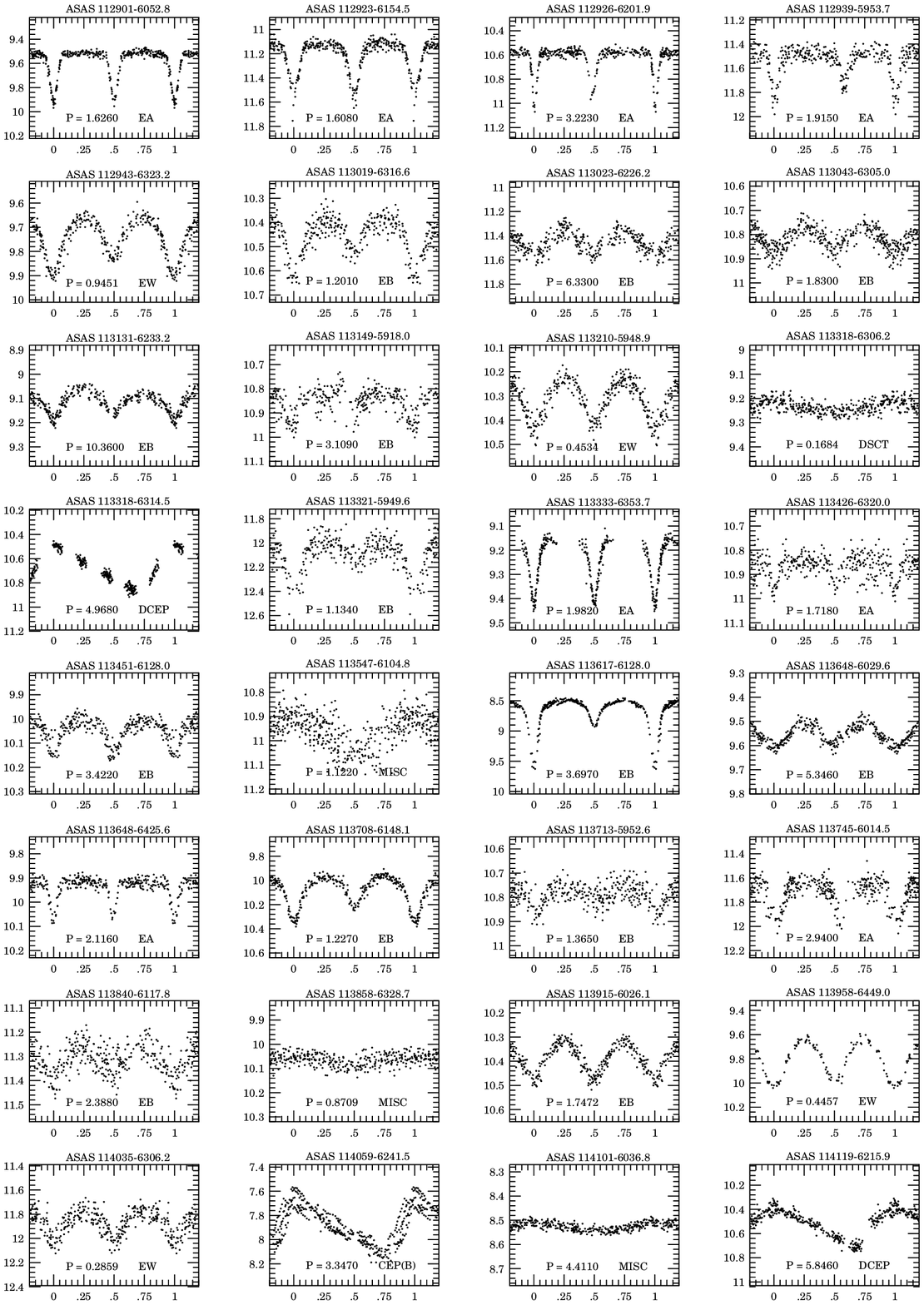}
\end{figure}
\begin{figure}\vspace*{29 cm}
\includegraphics{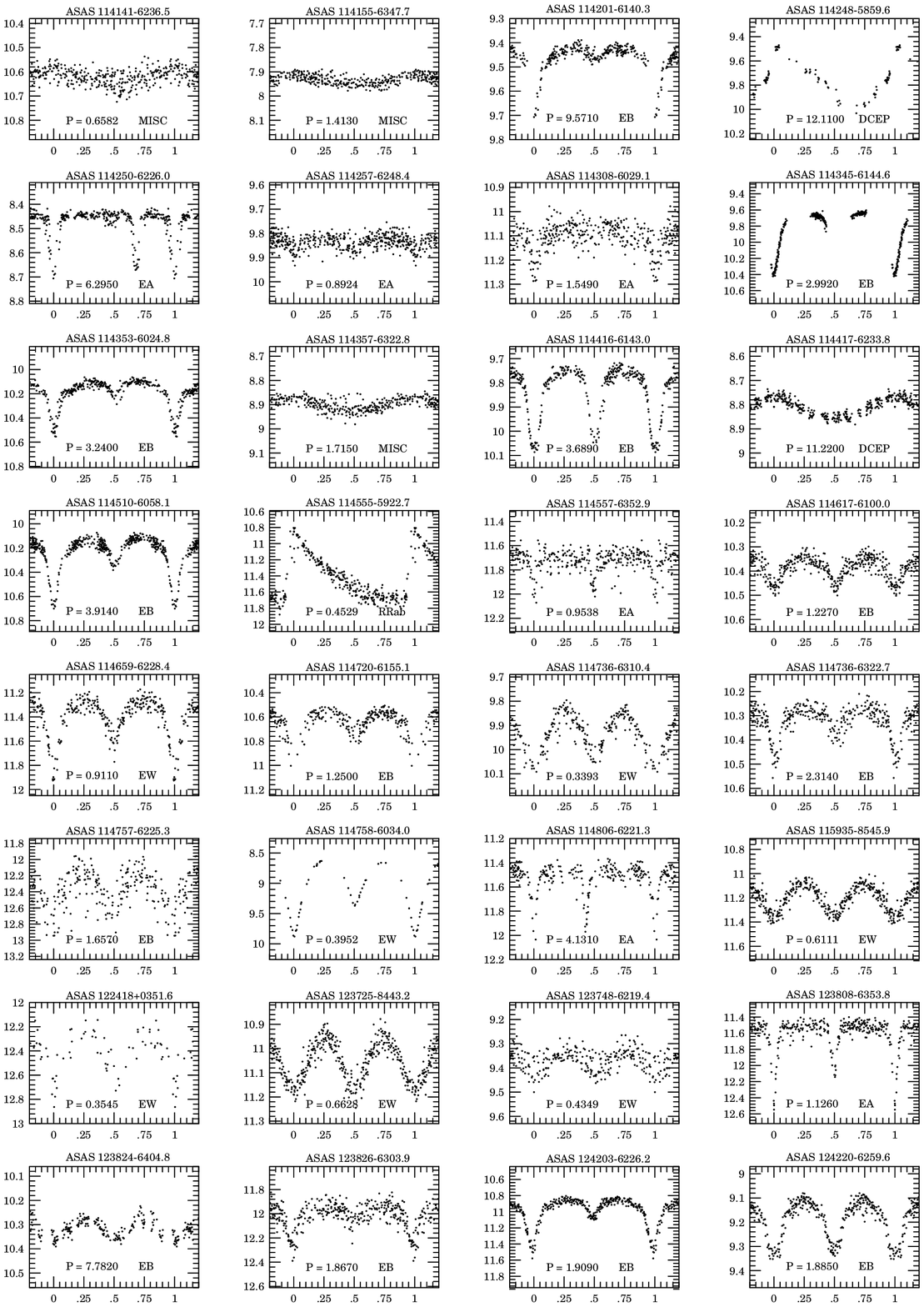}
\end{figure}
\begin{figure}\vspace*{29 cm}
\includegraphics{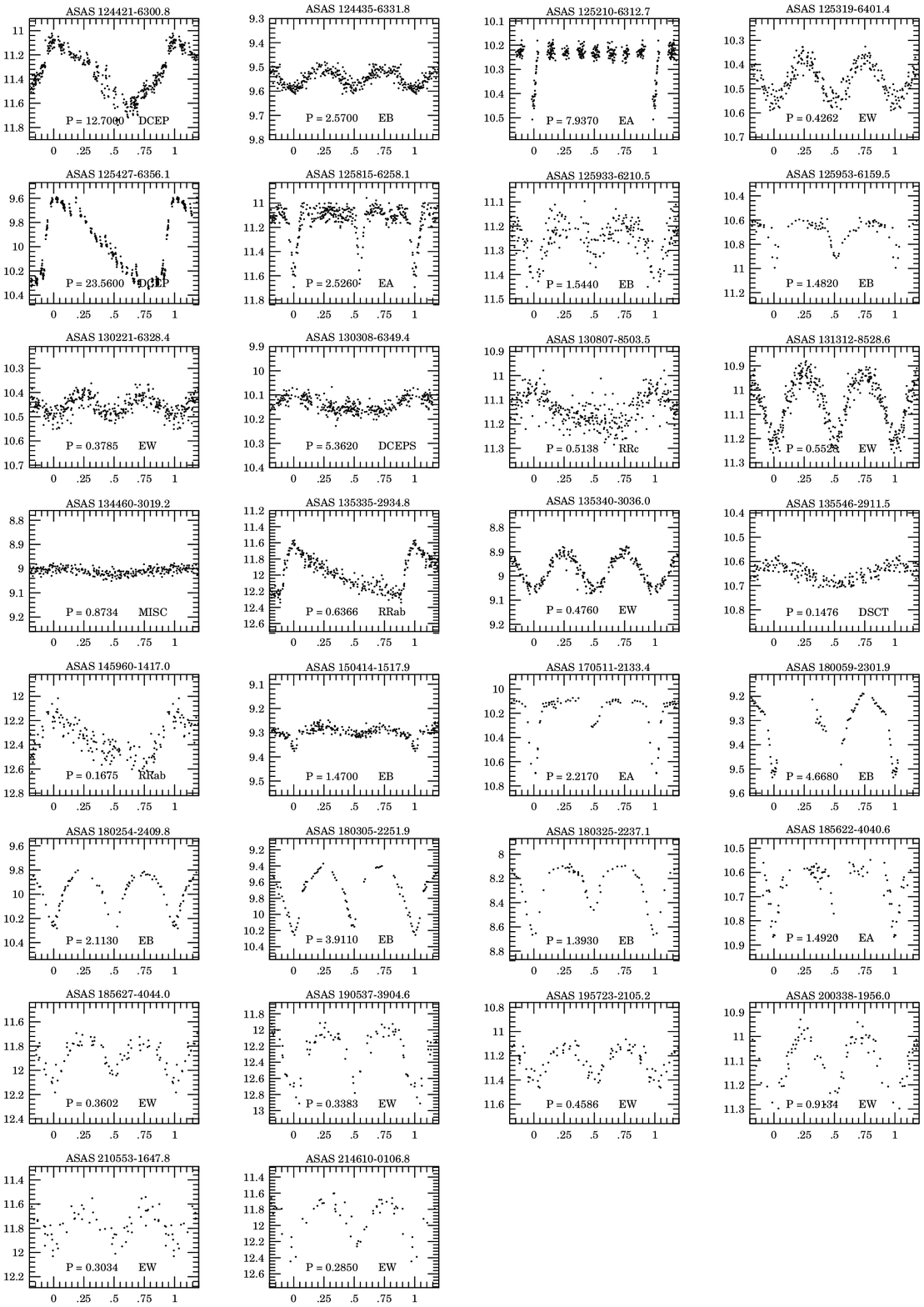}
\end{figure}
\end{document}